\documentclass[conference]{IEEEtran}
\IEEEoverridecommandlockouts
\usepackage{cite}
\usepackage{amsmath,amssymb,amsfonts}
\usepackage{algorithmic}
\usepackage{graphicx}
\usepackage{textcomp}
\usepackage{enumerate} % http://ctan.org/pkg/enumerate
\usepackage{booktabs}
\usepackage{adjustbox}
\usepackage{graphicx}
\usepackage{subcaption}
\usepackage{placeins} % \FloatBarrier forces float to be set without a pagebreak 
\usepackage{tikz}
\usepackage{listings} % \begin{lstlisting}[language=]
\usepackage{xcolor}
\def\BibTeX{{\rm B\kern-.05em{\sc i\kern-.025em b}\kern-.08em
		T\kern-.1667em\lower.7ex\hbox{E}\kern-.125emX}}
\colorlet{punct}{red!60!black}
\definecolor{background}{HTML}{EEEEEE}
\definecolor{delim}{RGB}{20,105,176}
\colorlet{numb}{magenta!60!black}

\lstdefinelanguage{json}{ 
	basicstyle=\normalfont\ttfamily,
	numbers=left,
	numberstyle=\scriptsize,
	stepnumber=1,
	numbersep=8pt,
	showstringspaces=false,
	breaklines=true,
	frame=lines,
	backgroundcolor=\color{background},
	literate=
	*{0}{{{\color{numb}0}}}{1}
	{1}{{{\color{numb}1}}}{1}
	{2}{{{\color{numb}2}}}{1}
	{3}{{{\color{numb}3}}}{1}
	{4}{{{\color{numb}4}}}{1}
	{5}{{{\color{numb}5}}}{1}
	{6}{{{\color{numb}6}}}{1}
	{7}{{{\color{numb}7}}}{1}
	{8}{{{\color{numb}8}}}{1}
	{9}{{{\color{numb}9}}}{1}
	{:}{{{\color{punct}{:}}}}{1}
	{,}{{{\color{punct}{,}}}}{1}
	{\{}{{{\color{delim}{\{}}}}{1}
	{\}}{{{\color{delim}{\}}}}}{1}
	{[}{{{\color{delim}{[}}}}{1}
	{]}{{{\color{delim}{]}}}}{1},
}
\usepackage[hidelinks]{hyperref} %makes references clickable and provides \autoref to include object type in reference %option hidelinks prevents ugly boxes in text
\usepackage{acronym}
\newacro{ttp}[TTPs]{Tactics, Techniques, and Procedures}
\newacro{lihp}[LIHP]{Low Interaction Honeypot}
\newacro{mihp}[MIHP]{Mid Interaction Honeypot}
\newacro{hihp}[HIHP]{High Interaction Honeypot}
\newacro{iot}[IoT]{Internet of Things}
\newacro{aws}[AWS]{Amazon Web Services}

\usepackage{cleveref} % provides cref and Cref to capitalize references
\crefname{subsection}{subsection}{subsections} %recognizes subsec label
\crefname{subsubsection}{subsubsection}{subsubsections} %%recognizes subsubsec label
\crefname{paragraph}{paragraph}{paragraphs} %recognizes paragraph label

\begin{document}

\title{A Qualitative Empirical Analysis of Human Post-Exploitation Behavior}
%{\footnotesize \textsuperscript{*}Note: Sub-titles are not captured in Xplore and should not be used}

%
% long author format
\author{
	\IEEEauthorblockN{Daniel Schneider\IEEEauthorrefmark{1}, Daniel Fraunholz\IEEEauthorrefmark{1}, Daniel Krohmer\IEEEauthorrefmark{1}} 
	\IEEEauthorblockA{\IEEEauthorrefmark{1}\textit{Intelligent Networks Research Group},
		\textit{German Research Center for Artificial Intelligence},
		D-67663 Kaiserslautern \\
		\{firstname\}.\{lastname\}@dfki.de}
	\thanks{This work has been supported by the German Federal Ministry of Education and Research (BMBF) (Foerderkennzeichen 01IS18062E, SCRATCh).
		The authors alone are responsible for the content of the paper.}
}

\maketitle

\begin{abstract}
Honeypots are a well-studied defensive measure in network security. This work proposes an effective low-cost honeypot that is easy to deploy and maintain.
The honeypot introduced in this work is able to handle commands in a non-standard way by blocking them or replying with an insult to the attacker.
To determine the most efficient defense strategy, the interaction between attacker and defender is modeled as a Bayesian two-player game. 
For the empirical analysis, three honeypot instances were deployed, each with a slight variation in its configuration.
In total, over 200 distinct sessions were captured, which allows for qualitative evaluation of post-exploitation behavior. The findings show that attackers react to insults and blocked commands in different ways, ranging from ignoring to sending insults themselves. The main contribution of this work lies in the proposed framework, which offers a low-cost alternative to more technically sophisticated and resource-intensive approaches. 
\end{abstract}


% body
\section{Introduction} \label{sec:intro}
\setcounter{page}{1}	% Set page numbering to begin on this page

Recent years have shown an increase in the quantity and severity of cyberattacks \cite{Ponemon.2019}. Although security concerns should therefore be addressed in every enterprise, the costs of doing so often seem too high, especially for small and mid-sized companies. This work proposes a security measure based on deception technology and game theory, which is inexpensive to implement, launch and maintain.

\subsection{Honeypots} \label{subsec:hon}
Honeypots are security resources which are unique in that they are specifically deployed so that malicious actors can interact with them, their sole goal is to be attacked \cite{Spitzner.2003}, which offers the advantage that all traffic observed on them can be regarded as malicious which greatly facilitates analysis \cite{Sqalli.2011}. 
Honeypots are used for mainly two reasons: Gathering information about \ac{ttp} of attackers (research), as well as luring attackers into spending time and resources on no-reward systems (production). Furthermore, honeypots can be classified according to their interaction level, i.e., whether they are only simulations of a system and only respond to a limited number of predefined commands (low interaction), or real operating systems which therefore support all the functions of a usual system (high interaction). \acp{lihp} are a lot cheaper to develop, deploy and maintain, but \acp{hihp} are considered to provide much more detailed information on attacks. \acp{mihp} combine aspects of both \acp{lihp} and \acp{hihp} \cite{Marin.2014}. 
A more general and in-depth overview of honeypot technology is given by \cite{Nawrocki.2016} and \cite{Fraunholz.2018a}.

\subsection{Game Theory} \label{subsec:game}
Game theory provides researches of many different backgrounds with the necessary tools to explore the process of decision-making \cite{Osborne.2006a}. Researchers of economics, sociology and political sciences all have been known to use game theory for a long time, and in recent years researchers of information security have come to employ the concepts to their own field as well \cite{Roy.2010}. A very early example of how game theory can be applied to the domain of deception was given by \cite{Spencer.1973}. More recently, \cite{Sallhammar.2007b} has developed increasingly complex game-theoretic models to help with the evaluation of a system's security. To assess the efficiency of defensive measures, \cite{Bistarelli.2007} combined defense trees and strategic game models, computing rewards based on economic indices. \cite{Garg.2007} applied extensive games toward a whole network of deceptive systems (honeynet). The addition of learning and self-adapting honeypot systems was incorporated into the game-theoretic literature by \cite{Wagener.2011b} and \cite{Pauna.2012}. Since honeypots have been considered as a defensive measure for networks consisting of \ac{iot} devices, \cite{La.2016} developed a Bayesian game model, in which both attacker and defender can behave deceptively. In this work, an original model is presented, thus adding to the literature at the intersection of game theory and information security. This introduces the second main research interest: The empirical evaluation of an original game theoretic model, which is explained in \Cref{sec:deception}.  

The rest of the paper is structured as follows: In order to demonstrate the application of game theory to network security, an original game theoretic model is established in \Cref{sec:deception}. Based on this model, an experiment is designed and carried out, which is described in \Cref{sec:exp}. The results of the experiment are discussed in \Cref{sec:disc}, and the work is concluded in \Cref{sec:con}. 

\section{Deception Game} \label{sec:deception}
In this section, a strategic situation between an attacker and a defender modeled with the help of game theory. \Cref{subsec:elements} identifies the elements of the game. In \Cref{subsec:rewards}, rewards for each player are assigned, and finally the game is solved by computing the Nash equilibrium in \Cref{subsec:equilbrium}. The actions which are available to the defender are informed by the work of \cite{Wagener.2011}, who configured four different actions: Allowing commands is the standard HIHP behavior, and also what is expected of a production system. Blocking commands and returning an error, or else substituting the server response for corrupted content, could lead to attackers trying out alternative commands, thus enhancing interaction count and attack duration. Lastly, insulting an attacker could provoke her to reply with insults, possibly revealing more information about herself, and additionally serve as a sort of reverse Turing test. 

\subsection{Game Elements} \label{subsec:elements}
The situation in which an attacker is confronted with a system that could be either a honeypot or a production system can be modeled as sequential Bayesian game, and shall be formalized as a tuple $(N, \Omega, A, p, P, \gamma)$, where $N = \{n_i : i \in \{1,2\}\}$ is the set of players such that player 1 is the attacker and player 2 is the defender. $\Omega = \{\omega_i : i \in \{1,2\}\}$ is the state set, with $\omega_1$ describing a deceptive defender and $\omega_2$ a non-deceptive defender. The selection of the defender's type is modeled as a move by nature. $A_A = \{a_{A,i} : i \in \{1,2\}\}$ is the action set for player 1 with $a_{A,1} = "attack"$ and $a_{A,2} = "resign"$. $A_D = \{a_{D,i} : i \in [1,3]\}$ is the action set of player 2, with $a_{D,1} = "allow"$, $a_{D,2} = "block"$, and $a_{D,3} = "insult"$. However, action $a_{D,3}$ is only available in state $\omega_1$, thus the action set $A_D$ is differentiated into $A_{D}^{\omega_1}$ and $A_{D}^{\omega_2}$. %\par 
In Bayesian games, it is not uncommon to establish a set $T$ which contains the signals that one or more player might send, often before the first action, in order to convince her opponent that she is of a certain type rather than the other. 
%For example, a player might want to signal that she is a "strong" opponent rather than a "weak" opponent. These signals can be separating, which means that only players of type x can rationally choose to send signal x, because for all other types, the cost would be too high. They can also be pooled, which means that players of all types can choose signal x, making it impossible to deduce a player's type with certainty based on her signals \cite{Osborne.2004a}. 
However in this case, the notion of observable actions is more conducive to a comprehensive description. It suggests that while one player may not be able to directly observe the other player's type, she can observe the other player's actions and form a belief about their type \cite{Osborne.2006c}. Note that the attacker does not know that $A_{D}^{\omega_1}$ and $A_{D}^{\omega_2}$ have different actions available, i.e. she cannot with certainty infer $\omega_1$ from observing $a_{D,3}$. %\par 
In the deception game presented here, player 1 initially assigns both states (separated by the defender's type) a certain probability based on a prior belief denoted by $p(\omega_1) = \frac{1}{10}$ and $p(\omega_2) = \frac{9}{10}$, meaning that the attacker estimates the probability of being confronted with a deceptive defender to be 10\%. During the game, the attacker's action-determined belief $P$ is continuously updated as shown in \Cref{bayes}. 
\begin{equation}\label{bayes}
P(\omega_i \mid a_i) = \frac{P(a_i \mid \omega_i) \cdot P(\omega_i)}{P(a_i)} 
\end{equation}
The reward function is given by $\gamma : N \times \Omega \times A \rightarrow \mathbf{Z}$.

\subsection{Rewards} \label{subsec:rewards}
In order to solve the model, each player's payoffs have to be determined. While a number of papers have used external factors to calculate payoffs \cite{Bistarelli.2007}, this is not a feasible approach here. Instead, the concept of subjective expected utility is used to arrive at the final rewards. The rewards discussed in this paragraph are formally described by the utility functions \Cref{rewarda}, \Cref{rewardnd} and \Cref{rewardd}. %\par 
For an attacker, every attack is associated with some cost $\mathbf{C}$ due to time and resources that have to be spent. This cost can be evaded by choosing to resign, but that ends the game and rules out any possible gains made possible by a successful attack. When faced with a non-deceptive defender, a successful attack provides the attacker with potentially valuable information or access $\mathbf{V}$, as well as emotional satisfaction $\mathbf{W}$ for reaching the goal and "winning" against the defender. In case of a deceptive defender, no material reward can be gained, and because this goal cannot be reached, the associated emotional satisfaction cannot be experienced. However, some satisfaction $\mathbf{S}$ may be derived from following intrinsic motivators such as a drive to explore systems, observe curious behavior or the enjoyment of the direct competition with an opponent. Additional reward stems from a more indirect competition for prestige $\mathbf{P}$ with other hackers. %\par
Attackers' motivations and interests can differ drastically from one another, which is why a sound attacker model should be established in real-life security endeavors \cite{Seebruck.2015}. Additionally, costs associated with launching an attack depend on both material and immaterial factors such as monetary cost for equipment, level of expertise and risk-aversion, which are nonuniform for different attackers \cite{Ingoldsby.2013}. 
\begin{equation} \label{rewarda}
\textnormal{SEU}_{A} = p(V) \cdot V + p(W) \cdot W + p(S) \cdot S + p(P) \cdot P - p(C) \cdot C
\end{equation}

A deceptive defender has a vested interest in being attacked and therefore aims to engage the attacker in order to gain as much insight $\mathbf{I}$ into attacker behavior as possible. Cost $\mathbf{C}$ arises through implausible behavior, which might deter potential attackers and thus minimize the chance to observe attack behavior. Note that in this case, cost is not fixed but weighted. This cost is partially offset by a utility $\mathbf{E}$, which rewards the defender for actions which create an environment that satisfies the attacker's intrinsic rewards $\mathbf{S}$, since those are fit to keep the attacker engaged. 
\begin{equation} \label{rewardd}
\textnormal{SEU}_{D}^{\omega_1} = p(I) \cdot I - (p(C) \cdot C - p(E) \cdot E)
\end{equation}

For a non-deceptive defender, the rewards are rather simple. Every security breach carries great cost $\mathbf{C}$, which is why malicious activity always has to be blocked as soon and as thoroughly as possible. Although the argument could be made that non-deceptive defenders might also gain potentially useful insight into attack patterns by allowing malicious traffic, this gain is minuscule when compared to the dangers of allowing a threat actor any kind of access to sensitive systems. 
\begin{equation} \label{rewardnd}
\textnormal{SEU}_{D}^{\omega_2} = - p(C) \cdot C
\end{equation}

% find equilibria
\subsection{Equilibrium} \label{subsec:equilbrium}
\begin{figure}
	\centering
	\begin{tikzpicture}[rotate=-90]
\tikzstyle{hollow}=[circle,draw,inner sep=1.5]
\tikzstyle{solid}=[circle,draw,inner sep=1.5,fill=black]
\tikzstyle{green}=[black!60!green]

\node[gray, hollow] (v1) at (0,0) {};
\node[gray]  (v2) at (2,2) {Deceptive};
\node[gray]  (v3) at (2,-2) {Non-deceptive};

\node[solid, red] (v4) at (4,2) {}; %1:1
\node[solid, red] (v15) at (4,-2) {}; %1:1

% 2:1
\node[solid, green] (v7) at (7,3) {}; %2:1
% attacker
\node[red] (v5) at (5.5,3) {Attack};
\node[red] (v6) at (5.5,1) {Resign};
% defender
\node[green] (v11) at (8.5,3) {Block};
\node[green] (v9) at (8.5,4) {Allow};
\node[green] (v13) at (8.5,2) {Insult};

%2:2
\node[solid, green] (v17) at (7,-1) {};
% attacker 
\node[red] (v16) at (5.5,-1) {Attack};
\node[red] (v18) at (5.5,-3) {Resign};
% defender
\node[green] (v20) at (8.5,-0.5) {Allow};
\node[green] (v22) at (8.5,-1.5) {Block};

% terminal nodes 
\node[solid, black] (v19) at (7.5,-3) {}; %1:1 nd resign
\node[solid, black] (v8) at (7.5,1) {}; %1:1 d resign
\node[solid, black] (v10) at (10,4) {}; %2:1 allow
\node[solid, black] (v12) at (10,3) {}; %2:1 block
\node[solid, black] (v14) at (10,2) {}; %2:1 insult
\node[solid, black] (v21) at (10,-0.5) {}; %2:2 allow
\node[solid, black] (v23) at (10,-1.5) {}; %2:2 block

% arrows
% chance move
\draw[gray]  (v1) edge (v2);
\draw[gray]  (v1) edge (v3);
\draw[gray]  (v2) edge (v4);
\draw[gray]  (v3) edge (v15);
% uncertainty
\draw[dashed, rounded corners=7, red]  (v4) edge (v15);

% deceptive
% attacker
\draw[red]  (v4) edge (v5);
\draw[red]  (v4) edge (v6);
\draw[red]  (v5) edge (v7);
\draw[red]  (v6) edge (v8);
% defender 
\draw[green]  (v7) edge (v9);
\draw[green]  (v9) edge (v10);
\draw[green]  (v7) edge (v11);
\draw[green]  (v11) edge (v12);
\draw[green]  (v7) edge (v13); 
\draw[green]  (v13) edge (v14);

%non-deceptive
% attacker
\draw[red]   (v15) edge (v16);
\draw[red]   (v16) edge (v17);
\draw[red]   (v15) edge (v18);
\draw[red]   (v18) edge (v19);
% defender
\draw[green]  (v17) edge (v20);
\draw[green]  (v20) edge (v21);
\draw[green]  (v17) edge (v22);
\draw[green]  (v22) edge (v23);

% seu d
\node[right, align=left, rotate=-90] at (10,4) {\tiny{$(SEU_{A} = p(V) \cdot V + p(W) \cdot W + p(S) \cdot S + p(P) \cdot P - p(C) \cdot C)$}, \\ \tiny{$(SEU_{D}^{\omega_1} = p(I) \cdot I - p(C) \cdot C)$}};
\node[right, align=left, rotate=-90] at (10,3) {\tiny{$(SEU_{A} = p(V) \cdot V + p(W) \cdot W + p(S) \cdot S + p(P) \cdot P - p(C) \cdot C)$}, \\ \tiny{$(SEU_{D}^{\omega_1} = p(I) \cdot I - p(C) \cdot C)$}};
\node[right, align=left, rotate=-90] at (10,2) {\tiny{$(SEU_{A} = p(V) \cdot V + p(W) \cdot W + p(S) \cdot S + p(P) \cdot P - p(C) \cdot C)$}, \\ \tiny{$(SEU_{D}^{\omega_1} = - p(C) \cdot C)$}};
\node[right, align=left, rotate=-90] at (7.5,1) {\tiny{$(SEU_{A} = p(V) \cdot V + p(W) \cdot W + p(S) \cdot S + p(P) \cdot P - p(C) \cdot C)$}, \\ \tiny{$(SEU_{D}^{\omega_1} = p(I) \cdot I - p(C) \cdot C)$}};

% seu nd
\node[right, align=left, rotate=-90] at (10,-0.5) {\tiny{$(SEU_{A} = p(V) \cdot V + p(W) \cdot W + p(S) \cdot S + p(P) \cdot P - p(C) \cdot C)$}, \\ \tiny{$(SEU_{D}^{\omega_2} = - (p(C) \cdot C - p(E) \cdot E))$}};
\node[right, align=left, rotate=-90] at (10,-1.5) {\tiny{$(SEU_{A} = p(V) \cdot V + p(W) \cdot W + p(S) \cdot S + p(P) \cdot P - p(C) \cdot C)$}, \\ \tiny{$(SEU_{D}^{\omega_2} = - (p(C) \cdot C - p(E) \cdot E))$}};
\node[right, align=left, rotate=-90] at (7.5,-3) {\tiny{$(SEU_{A} = p(V) \cdot V + p(W) \cdot W + p(S) \cdot S + p(P) \cdot P - p(C) \cdot C)$}, \\ \tiny{$(SEU_{D}^{\omega_2} = - (p(C) \cdot C - p(E) \cdot E))$}};
\end{tikzpicture}
	\caption{Game tree, style conventions as seen in \cite{Osborne.2006c}}
	\label{fig:game_tree}
\end{figure}
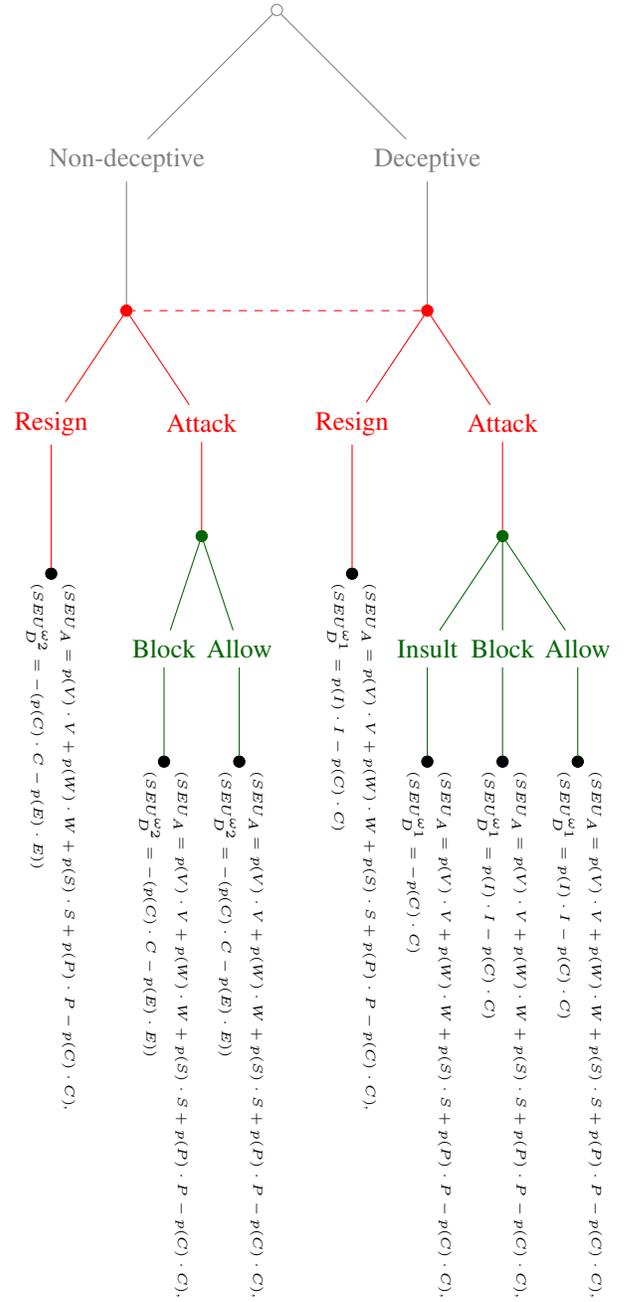

After the outcome values have been assigned by constructing SEU functions, the game still has to be solved i.e. the equilibrium has to be determined, for which integer values are approximated for each variable in the respective SEU functions, thus resulting in easily comparable numerical values. 
First, nodes are eliminated iteratively by checking for domination. %\cite{Osborne.2006b}. 
This reveals that $(a_{A,1}^{\omega_2},a_{D,2}^{\omega_2})$ is strictly dominated at the first iteration. At the second iteration, $a_{A2}$ is strictly dominated in both $\omega_1$ and $\omega_2$. Second, the equilibrium is calculated by making use of the algorithm developed by \cite{Turocy.2010a}, which is based on the work of \cite{McKelvey.1995} who leverage quantal response equilibria.
% in a way that is conceptually similar to the Tracing Procedure introduced by \cite{Harsanyi.1975}. 
For games in normal form rather than extensive form, an algorithm was proposed by \cite{Turocy.2005}. The algorithm is used via its implementation in \textit{Gambit} \cite{McKelvey.2014}. This results in exactly one Nash equilibrium where the attacker has a pure strategy of choosing $a_{A,1}$ to attack, while the defender has a mixed strategy in which all three actions are played with a probability of $\frac{1}{3}$ in $\omega_1$. In $\omega_2$, the defender always chooses $a_{D,2}$.  The resulting game tree is depicted in \Cref{fig:game_tree}, where the moves by nature are gray, the attacker's moves are red and the defender's moves are colored green. The terminal nodes are black. To symbolize the attacker's uncertainty about the defender's type, the information sets are connected by a dashed line.

\section{Experiment} \label{sec:exp}
%The following section describes the orginal field experiment which serves to provide empirical data for this work. First, the purpose of the experiment is described in \Cref{subsec:purp}. \Cref{subsec:imp} shows the technical implementation of the needed aspects. The general setup of the experiment is described in \Cref{subsec:setup}. \Cref{subsec:data} presents an overview of the data generated during the course of the experiment. The analytical methods used to analyze this data are presented in \Cref{subsec:meth}. Finally, \Cref{subsec:res} shows the results of the data analysis. 

\subsection{Purpose} \label{subsec:purp}
The purpose of the original field experiment described in this section is twofold.
Firstly, this experiment is designed to provide empirical data to test how well the model established in \Cref{sec:deception} is applicable to real-life scenarios. This is done in accordance with an interpretation of game theory not as science of how games should optimally be played by ideal players with unlimited cognitive resources, but as tool to further the understanding of how actual people behave in certain situations.
Secondly, the experiment is also intended to compare the performance of different honeypot configurations. As mentioned in \Cref{subsec:hon}, there is a vast range concerning honeypots when it comes to complexity and sophistication. The honeypot based on the model established in \Cref{sec:deception} is compared to an almost identical honeypot, varying only in one aspect of the configuration: The probabilities for each action are derived from the results of \cite{Wagener.2011}, who used sophisticated technical solutions to optimize honeypot performance. The comparison is made based upon the number of an attacker's interactions with the honeypot, as well as the duration of an attack. 
One main innovation in the work of \cite{Wagener.2011} was the introduction of insults into honeypot systems. The reasoning was that human attackers might be engaged by verbal attacks, and in their response reveal more information about themselves that might be useful for forensic analysis. In their empirical test, \cite{Wagener.2011} were able to show that this approach can be fruitful. This work is trying to replicate those findings in a different setting.

\subsection{Implementation} \label{subsec:imp}
Over the course of the experiment, three different honeypots were launched, all of whom are based on \textit{Cowrie} \cite{Oosterhof.2018} which is a MIHP written in \textit{Python}. %\cite{Python.2}. 
The default way this honeypot handles command line input is to check the input against a list of commands, which emulate common bash commands, and execute the respective function in case of a successful check. The first honeypot, named \textit{Gamepot} behaves as concluded in \Cref{sec:deception}, meaning that it chooses each action ($"allow", "block", "insult"$) with a respective probability of 33\%. This probabilistic behavior has been achieved by modifying the module in \verb|cowrie-dev/src/cowrie/shell/honeypot.py| via the \texttt{random.choice()} method. Both added functions handle command line input by circumventing the default check and replying with either an error message taken from the list output by \texttt{errno --list}, or an insult taken from the list used by \texttt{sudo} when the insult function is activated. Modifications were made in the same module to properly reflect this behavior in the log files automatically created by \textit{Cowrie}. The second honeypot was configured in such a way that the probabilities for each action reflect the values presented in \cite{Wagener.2011} as the result of an adaptive process based on machine learning. As a tribute to their work, this honeypot is called \textit{Heliza}, since it essentially tries to emulate their configuration as a MIHP. The third and last honeypot served as a control instance, allowing all possible commands, and is hence named \textit{Control}. 

\subsection{Setup} \label{subsec:setup}
Apart from the differences described in \Cref{subsec:imp}, all three honeypots were set up identically to control as many variables as possible. Two users were created, each with his own home directory. Both home directories had directories "work" and "private", which were populated with non-sensitive directories and files. This was done to provide a plausible file structure for attackers to explore, which awards two benefits: First, research has shown that human attackers spend more time on systems with greater file system depth \cite{Farinholt.2017}, which is one of the core interests of honeypot operations. Second, a seemingly authentic file system may help to ease an attacker's suspicion of interacting with a honeypot.
All three honeypots have been hosted by \ac{aws}, which comes with several advantages. First, even though it is highly unlikely that an attacker could use the environment provided by \textit{Cowrie} to do any actual harm, it is always preferable to host research honeypots outside of one's own network for security reasons. Second, AWS offers virtual server instances in many locations, including the US, which has been shown to be more popular with attackers than other regions. Additionally, servers hosted by AWS tend to attract high numbers of attackers in a short amount of time \cite{Barron.2017}.
Both \textit{Heliza} and \textit{Control} went online on October 5th, 2018. At this point, \textit{Gamepot} was had not received its final configuration yet, which is why it had to be launched later, on December 21st, 2018. Possible issues regarding the unequal uptime of the three instances are discussed in \Cref{sec:disc}.  
After bringing a new instance online, it only takes mere minutes until the first attack. However, those are usually automated attacks, often as part of botnet operations, which pose a massive threat to network security \cite{Hoque.2015,Kolias.2017}. Since this work is interested in the non-automated behavior of human attackers, measures had to be taken to discourage attacks by bots and simultaneously invite attacks by humans. To mitigate brute force attacks, strong passwords were chosen for the required login credentials. Additionally, \textit{Fail2Ban} \cite{Jaquier.2018} was employed to blacklist IP addresses after a certain number of failed login attempts inside a specified time frame. To attract human attackers, rather than just keeping bots out, the correct login credentials were automated to be leaked on paste sites, since this approach has been shown to be of low cost, yet effective \cite{Fraunholz.2018c}. 

\subsection{Data} \label{subsec:data}
The gathered data, which has a volume of over three gigabytes in its raw format, consists of log files detailing every interaction with the honeypot, which are automatically created by \textit{Cowrie} and stored in a newline-delimited \textit{JavaScript Object Notation} (JSON)  format. Each JSON object inside a log file contains a unique session ID with which attack sessions can be distinguished, along with identifying data like IP addresses and timestamps. 
%A non-sensitive example of such a JSON object is shown below. It documents the initialization of a new session during the testing phase on a secure internal network. 

%%\newpage
%\begin{lstlisting}[language=json]
%{
%"eventid": "cowrie.session.connect", 
%"src_ip": "192.168.53.210", 
%"src_port": 50689, 
%"timestamp": "2018-08-13T14:12:32.585156Z", 
%"message": "New connection: 192.168.53.210:50689 (192.168.53.221:2222) [session: 876c600fcbda]",
%"dst_ip": "192.168.53.221", 
%"protocol": "ssh", 
%"session": "876c600fcbda", 
%"dst_port": 2222, 
%"sensor": "debian"
%}
%\end{lstlisting}

\subsection{Methods} \label{subsec:meth}
The raw data has been parsed using \textit{Python}.  
Calculation as visualization of descriptive statistics has been done in \textit{R}, a programming language for statistical computation.

\subsection{Quantitative Results} \label{subsec:res}
One of the key interests of the experiment was to show if the configuration described in \Cref{subsec:imp} is more effective in engaging attackers compared to other configurations. As measurement, both attack duration (time spent on the honeypot by each attacker) and interaction quantity (number of commands issued by the attacker during a session) are evaluated in this section. 
A brief summary statistic of the attack duration varying between the three different configurations is presented in \Cref{tab:duration}.
It shows that the \textit{Gamepot} instance, based on 14 observations, has a lower average attack duration in comparison the both the \textit{Heliza} instance, with 67 observations, and the \textit{Control} group with 149 observations. The instance with the highest average attack duration is \textit{Control}, with \textit{Heliza} and \textit{Control} having an almost identical interquartile range. This result is unexpected, since it not only does not show any advantage of \textit{Gamepot} over \textit{Heliza}, it also shows no advantage of either of them over the control group. This contradiction is further discussed in \Cref{sec:disc}. All three instances show a maximum duration of around 900 seconds, due to \textit{Cowrie} setting a nonrigid cap at this mark.

\begin{table}[!htbp] 
	\centering 
	\caption{Attack duration in seconds, inter-instance comparison} 
	\label{tab:duration} 
	\begin{tabular}{lccccccc} 
		\toprule
		Instance & \multicolumn{1}{c}{N} & \multicolumn{1}{c}{Min.} & \multicolumn{1}{c}{1st Qu.} & \multicolumn{1}{c}{Median} & \multicolumn{1}{c}{Mean} & \multicolumn{1}{c}{3rd Qu.} & \multicolumn{1}{c}{Max} \\ 
		\midrule
		Gamepot & 14 & 5.314 & 80.093 & 118.787 & 329.870 & 591.122 & 902.315 \\
		Heliza & 67 & 10.07 & 67.44 & 170.47 & 419.69 & 898.93 & 942.60 \\
		Control & 149 & 3.582 & 66.912 & 234.876 & 422.611 & 898.740 & 933.530 \\  
		\bottomrule
	\end{tabular} 
\end{table} 

To compare the different configurations with regard to how much attackers interacted with each honeypot, attack sessions on each instance have been grouped in inter-instantially uniform bins based on observed clustering. The first bin consists of those sessions where attackers successfully logged in, but did not issue any commands. The last bin has a vast range, from 350 to 700 commands, to catch outliers on the top end.  

\Cref{tab:interaction} presents a metric overview of attack durations, which includes both total and percentage values. In combination, it becomes apparent that on the \textit{Gamepot} instance, a lot less attackers logged in without following up with any commands, relatively. However, due to the very small number of observations, this can be only the most timid of hints toward a generalizable tendency. The fact that the cluster with zero commands is so strongly represented conflicts with the expectations of the game theoretic model in \Cref{subsec:equilbrium}, which is discussed in \Cref{sec:disc}. One other difference worth mentioning is that the control group is relatively lacking in the high clusters, which might point to this configuration being a less immersive environment.

\begin{table*}
	\centering
	\caption{Interaction in number of commands} 
	\label{tab:interaction} 
	\scriptsize
	\begin{tabular}{lccccccccc} 
		\toprule
		Instance & & \multicolumn{1}{c}{[0,1]} & \multicolumn{1}{c}{(1,15]} & \multicolumn{1}{c}{(15,35]} & \multicolumn{1}{c}{(35,50]} & \multicolumn{1}{c}{(50,100]} & \multicolumn{1}{c}{(100,200]} & \multicolumn{1}{c}{(200,350]} & \multicolumn{1}{c}{(350,700]} \\ 
		\midrule 
		Gamepot & total & 1 & 4 & 3 & 1 & 3 & 0 & 2 & 0 \\
		& \% & 7.14 & 28.57 & 21.43 & 7.14 & 21.43 & 0.00 & 14.29 & 0.00  \\
		Heliza & total & 15 & 15 & 10 & 5 & 6 & 5 & 9 & 2 \\
		& \% & 22.39 & 22.39 & 14.93 & 7.46 & 8.96 & 7.46 & 13.43 & 2.99  \\
		Control & total & 50 & 28 & 18 & 19 & 21 & 10 & 3 & 0 \\  
		& \% & 33,56 & 18.79 & 12.08 & 12.7 & 14.09 & 6.71 & 2.01 & 0.00  \\
		\bottomrule 
	\end{tabular}
\end{table*}

In sum, the experiment presented here was not able to perform as expected.
Possible reasons and conceivable countermeasures for this result are discussed in \Cref{sec:disc}. 

\subsection{Qualitative Results} \label{subsec:qual}
Exploiting the handy number of attack sessions, a qualitative analysis into attack patterns and behavior can be done. What follows is a more in-depth view of a particularly interesting set of attack sessions, originating from the same IP address and using the same client-side user agent, was observed on the helzia honeypot. This set consists of nine individually captured sessions.
Although this is not definite proof, it's likely that the person on the attacker-side of these sessions is the same.
However, some observations vary vastly between the sessions. In the first sessions,
the trespasser tried to install a cryptomining software called luk miner from dropbox, trying out several different versions.
As the attacker tried to start the software, the intended mining pool (pool.simpleco.in) as well as username (utkarshkg.20762) and password (x) could be obtained by the authors.
The session was immediately ended after the attacker noticed that the downloaded software is unable to run on the system.
Interestingly in this set, the response to blocked actions as well as to insults is always to just repeat the command.
This suggests that the information obtained by the additional commands are not useful from a defenders perspective, even though common metrics such as number of issued commands or duration of session would indicate a higher information gain.
In rare cases, attackers reacted to insults or blocks by simply hitting the return key multiple times, without entering anything into the command line. While this kind of behavior is more typically identified as bot activity, the high number of occurrences of typos suggest that the honeypots interacted with a human being.
The second session of this individual equals the first, except that instead of dropbox, the official luk miner sources (www.lukminer.net) are used to download the software.
After the second session the behavior changes drastically, several attempts to compromise the root account via \texttt{sudo} and the sudoers file were observed followed by several attempts to update the system.
After these attempts being unsuccessful, the attacker changed to destructive behavior, trying to destroy the system by issuing \texttt{rm -rf /}, which would render a system unusable in case sufficient privileges were available.
Directly after those attempts were deemed unsuccessful as well, the attacker tried to install different desktop environments (\textit{xfce}) and server-side software to forward those (\textit{xrdp} and \textit{vnc}).
From there, the attacker appears to suspect the network configuration to be responsible for the suspicious behavior. First, firewall software (\textit{ufw}) and later communication software (\textit{firefox}, \textit{chrome} and \textit{ssh}) were tried to be installed. In one observed session, the attacker tried to fortify the honeypot by deploying firewall software and issuing the commands to block any incoming request except for ssh access, which is the protocol the attacker used to communicate with the system.
However, the described issue with the actual information gain from additional commands remains unchanged in all observed sessions from this source.
For the purpose of gaining insight into the attacker's tactics and tools, the honeypot's capability to save all downloaded content in a directory removed from the attacker's reach, enabled in the default configuration of \textit{Cowrie}, appeared to be more effective than the insult and block functionalities implemented by the authors, as attackers tended to download software from an additional source or try another exploitation method if the first failed, while blocking and insulting merely resulted in a repetition of the previous command in most cases. 
In fact, in no observed session an actual information gain could archived by blocking or insulting, i.e. even in cases where attackers did not repeat the rejected command but rather tried a different command, no additional repositories or addresses were revealed. However, taking into account that wasting an attacker's time and resources is also a common objective for honeypot operations, the added functionalities have shown some success. 

\section{Discussion} \label{sec:disc}
There are several issues and limitations regarding this work which have to be discussed. One issue is that of the unequal length of observation phases for the three honeypots, which naturally results in unequal numbers of observed sessions, making a direct comparison between the instances rather difficult, from a quantitative perspective. In a similar manner, the fact that the \textit{Gamepot} honeypot has such a small observation count results in huge error margins that do not allow for proper statistical analysis. 
A second issue lies in the sizable number of attackers who managed to log in, but did not follow up with any commands. This is inconsistent with what would be expected of an attacker in light of the game theoretic model established in \Cref{sec:deception}. A possible reason for this might be that the model underestimates the attacker's prior belief to be confronted with a deceptive defender, although in this case it is unclear why an attacker would even try to log in. Another possibility is that many attackers in this sample are driven by a non-malicious curiosity, that was satisfied upon realizing that the leaked credentials actually do work.   
%
%Reflecting on the unexpected results concerning the \textit{Control} instance, there is some doubt about whether the control group was conceptualized correctly. Although it does make sense to compare the performance of a honeypot with standard behavior to such of non-standard behavior, a different approach to constructing the control group might have possibly been more fruitful in this context. 
%
The potentially biggest concern is that of the observed attack duration. While a result showing that the \textit{Heliza} instance has a higher average attack duration than the \textit{Gamepot} instance would simply point towards a superior performance of the former, the fact that the \textit{Control} instance has the highest average attack duration is perplexing. Naturally, this observation might be caused simply by the extremely narrow number of cases. However, if this result were to be replicated in larger experiments, it would point to a conceptional failure of such non-standard behaviors in honeypot configurations. 
%
%A general issue that this study could be charged with lies in criticism leveled at the rational agent paradigm both generally \cite{Osborne.2006a} and more specificall in the context of online delinquency, namely that framing online delinquents as rational agents on a quest for maximum gain would not depict reality \cite{Garg.2015}. However, this kind of criticism has been addressed almost three decades ago by \cite{Esser.1990}, who stated that concepts are most fruitfully discussed and improved upon by applying them to concrete cases, which is what has been sought to accomplish in this study.  

\section{Conclusion} \label{sec:con}
In this work, an original game theoretic model was developed, modeling the confrontation of an attacker with a system, where the attacker is unsure if the system is a honeypot or a production system. From this mode, the configuration of a medium interaction honeypot was inferred and implemented. The main idea was to test a specific approach, i.e. enabling a honeypot to block an attacker's command and returning either an error message or an insult. To evaluate the effectiveness of this approach, an empirical study was performed, with two honeypots showing different variations of the feature and one remaining in default configuration, serving as control group. Although the presented experiment could not provide clean results, the general approach might still serve as a starting point for further research in this exciting intersection of academic research. Future endeavors in this direction should run for an enlarged period of time, as to achieve a higher number of observable attack sessions. 

%
%
% ---- Bibliography ----
%
%
\bibliographystyle{IEEEtran}
\bibliography{ngna2020}

% Generated by IEEEtran.bst, version: 1.14 (2015/08/26)
\begin{thebibliography}{10}
\providecommand{\url}[1]{#1}
\csname url@samestyle\endcsname
\providecommand{\newblock}{\relax}
\providecommand{\bibinfo}[2]{#2}
\providecommand{\BIBentrySTDinterwordspacing}{\spaceskip=0pt\relax}
\providecommand{\BIBentryALTinterwordstretchfactor}{4}
\providecommand{\BIBentryALTinterwordspacing}{\spaceskip=\fontdimen2\font plus
\BIBentryALTinterwordstretchfactor\fontdimen3\font minus
  \fontdimen4\font\relax}
\providecommand{\BIBforeignlanguage}[2]{{%
\expandafter\ifx\csname l@#1\endcsname\relax
\typeout{** WARNING: IEEEtran.bst: No hyphenation pattern has been}%
\typeout{** loaded for the language `#1'. Using the pattern for}%
\typeout{** the default language instead.}%
\else
\language=\csname l@#1\endcsname
\fi
#2}}
\providecommand{\BIBdecl}{\relax}
\BIBdecl

\bibitem{Ponemon.2019}
K.~Bissell, R.~M. Lasalle, and P.~{Dal Cin}, ``Ninth annual cost of cybercrime
  study,''
  https://www.accenture.com/us-en/insights/security/cost-cybercrime-study,
  2019.

\bibitem{Spitzner.2003}
L.~Spitzner, ``Honeypots: Catching the insider threat,'' in \emph{Computer
  Security Applications Conference, 2003. Proceedings. 19th Annual}, 2003, pp.
  170--179.

\bibitem{Sqalli.2011}
M.~H. Sqalli, S.~N. Firdous, Z.~Baig, and F.~Azzedin, ``An entropy and
  volume-based approach for identifying malicious activities in honeynet
  traffic,'' in \emph{Cyberworlds (CW), 2011 International Conference
  on}.\hskip 1em plus 0.5em minus 0.4em\relax IEEE, 2011, pp. 23--30.

\bibitem{Marin.2014}
\BIBentryALTinterwordspacing
J.~M.~F. Mar{\'i}n, J.~{\'A}.~M. Naranjo, and L.~G. Casado, \emph{Honeypots and
  Honeynets: Analysis and Case Study}.\hskip 1em plus 0.5em minus 0.4em\relax
  {IGI Global}, 2014, pp. 452--482. [Online]. Available:
  \url{https://www.igi-global.com/ViewTitle.aspx?TitleId=115776}
\BIBentrySTDinterwordspacing

\bibitem{Nawrocki.2016}
M.~Nawrocki, M.~W{\"a}hlisch, T.~C. Schmidt, C.~Keil, and J.~Sch{\"o}nfelder,
  ``A survey on honeypot software and data analysis,'' \emph{arXiv preprint
  arXiv:1608.06249}, 2016.

\bibitem{Fraunholz.2018a}
D.~Fraunholz, S.~{Duque Anton}, C.~Lipps, D.~Reti, D.~Krohmer, F.~Pohl,
  M.~Tammen, and H.~D. Schotten, ``Demystifying deception technology: A
  survey,'' \emph{arXiv preprint arXiv:1804.06196}, 2018.

\bibitem{Osborne.2006a}
M.~J. Osborne and A.~Rubinstein, \emph{A course in game theory}, 12th~ed.\hskip
  1em plus 0.5em minus 0.4em\relax Cambridge, Mass.: {MIT Press}, 2006, ch.~1.

\bibitem{Roy.2010}
S.~Roy, C.~Ellis, S.~Shiva, D.~Dasgupta, V.~Shandilya, and Q.~Wu, ``A survey of
  game theory as applied to network security,'' in \emph{2010 43rd Hawaii
  International Conference on System Sciences}.\hskip 1em plus 0.5em minus
  0.4em\relax {IEEE}, 2010.

\bibitem{Spencer.1973}
J.~Spencer, ``A deception game,'' \emph{American Mathematical Monthly},
  vol.~80, pp. 416--417, 1973.

\bibitem{Sallhammar.2007b}
K.~Sallhammar, ``Stochastic models for combined security and dependability
  evaluation,'' Thesis for the degree of philosophiae doctor, {Norwegian
  University of Science and Technology}, Trondheim, 2007.

\bibitem{Bistarelli.2007}
S.~Bistarelli, M.~Dall'Aglio, and P.~Peretti, ``Strategic games on defense
  trees,'' in \emph{Formal aspects in security and trust}, ser. Lecture Notes
  in Computer Science, T.~Dimitrakos, F.~Martinelli, P.~Y.~A. Ryan, and
  S.~Schneider, Eds.\hskip 1em plus 0.5em minus 0.4em\relax Berlin: Springer,
  2007, pp. 1--15.

\bibitem{Garg.2007}
N.~Garg and D.~Grosu, ``Deception in honeynets: A game-theoretic analysis,'' in
  \emph{Annual IEEE SMC Information Assurance and Security Workshop (IAW)},
  2007.

\bibitem{Wagener.2011b}
G.~Wagener, ``Self-adaptive honeypots coercing and assessing attacker
  behaviour,'' Ph.D. dissertation, {Institut National Polytechnique de Lorraine
  - INPL}, 2011.

\bibitem{Pauna.2012}
A.~Pauna, ``Improved self adaptive honeypots capable of detecting rootkit
  malware,'' in \emph{2012 9th International Conference on Communications
  (COMM)}.\hskip 1em plus 0.5em minus 0.4em\relax IEEE, 2012, pp. 281--284.

\bibitem{La.2016}
Q.~D. La, T.~Q.~S. Quek, and J.~Lee, ``A game theoretic model for enabling
  honeypots in iot networks,'' in \emph{2016 IEEE International Conference on
  Communications (ICC)}.\hskip 1em plus 0.5em minus 0.4em\relax IEEE, 2016, pp.
  1--6.

\bibitem{Wagener.2011}
G.~Wagener, R.~State, A.~Dulaunoy, and T.~Engel, ``Heliza: talking dirty to the
  attackers,'' \emph{Journal in Computer Virology}, vol.~7, no.~3, pp.
  221--232, 2011.

\bibitem{Osborne.2006c}
M.~J. Osborne and A.~Rubinstein, \emph{A course in game theory}, 12th~ed.\hskip
  1em plus 0.5em minus 0.4em\relax Cambridge, Mass.: {MIT Press}, 2006, ch.~12.

\bibitem{Seebruck.2015}
\BIBentryALTinterwordspacing
R.~Seebruck, ``A typology of hackers: Classifying cyber malfeasance using a
  weighted arc circumplex model,'' \emph{Digital Investigation}, vol.~14, pp.
  36--45, 2015. [Online]. Available:
  \url{https://doi.org/10.1016/j.diin.2015.07.002}
\BIBentrySTDinterwordspacing

\bibitem{Ingoldsby.2013}
T.~R. Ingoldsby, ``Attack tree-based threat risk analysis,'' \emph{Amenaza
  Technologies Limited}, vol. 550, 2013.

\bibitem{Turocy.2010a}
T.~L. Turocy, ``Computing sequential equilibria using agent quantal response
  equilibria,'' \emph{Economic Theory}, vol.~42, no.~1, pp. 255--269, Jan 2010.

\bibitem{McKelvey.1995}
R.~D. McKelvey and T.~R. Palfrey, ``Quantal response equilibria for normal form
  games,'' \emph{Games and economic behavior}, vol.~10, no.~1, pp. 6--38, 1995.

\bibitem{Turocy.2005}
T.~L. Turocy, ``A dynamic homotopy interpretation of the logistic quantal
  response equilibrium correspondence,'' \emph{Games and Economic Behavior},
  vol.~51, no.~2, pp. 243 -- 263, 2005.

\bibitem{McKelvey.2014}
R.~D. McKelvey, A.~M. McLennan, and T.~L. Turocy, ``Gambit: Software tools for
  game theory,'' http://www.gambit-project.org, 2014, version 13.1.2.

\bibitem{Oosterhof.2018}
\BIBentryALTinterwordspacing
M.~Oosterhof, ``Cowrie: Ssh/telnet honeypot,'' 2018. [Online]. Available:
  \url{https://github.com/micheloosterhof/cowrie}
\BIBentrySTDinterwordspacing

\bibitem{Farinholt.2017}
B.~Farinholt, M.~Rezaeirad, P.~Pearce, H.~Dharmdasani, H.~Yin, S.~{Le Blond},
  D.~McCoy, and K.~Levchenko, ``To catch a ratter: Monitoring the behavior of
  amateur darkcomet rat operators in the wild,'' \emph{Proceedings of the 38th
  IEEE Symposium on~Security and Privacy}, vol.~38, no.~1, 2017.

\bibitem{Barron.2017}
T.~Barron and N.~Nikiforakis, ``Picky attackers: Quantifying the role of system
  properties on intruder behavior,'' in \emph{Proceedings of the 33rd Annual
  Computer Security Applications Conference on - ACSAC 2017}.\hskip 1em plus
  0.5em minus 0.4em\relax New York, New York, USA: {ACM Press}, 2017, pp.
  387--398.

\bibitem{Hoque.2015}
N.~Hoque, D.~K. Bhattacharyya, and J.~K. Kalita, ``Botnet in ddos attacks:
  Trends and challenges.'' \emph{IEEE Communications Surveys and Tutorials},
  vol.~17, no.~4, pp. 2242--2270, 2015.

\bibitem{Kolias.2017}
C.~Kolias, G.~Kambourakis, A.~Stavrou, and J.~Voas, ``Ddos in the iot: Mirai
  and other botnets,'' \emph{Computer}, vol.~50, no.~7, pp. 80--84, 2017.

\bibitem{Jaquier.2018}
C.~Jaquier, Y.~Halchenko, D.~Black, S.~Hiscocks, and A.~Busleiman,
  ``Fail2ban,'' https://github.com/fail2ban/fail2ban, 2018, version 0.10.3.1.

\bibitem{Fraunholz.2018c}
D.~Fraunholz, D.~Schneider, J.~Zemitis, and H.~D. Schotten, ``Hack my company:
  An empirical assessment of post-exploitation behavior and lateral movement in
  cloud environments,'' in \emph{Proceedings of the Central European
  Cybersecurity Conference 2018}, Central European Cybersecurity
  Conference.\hskip 1em plus 0.5em minus 0.4em\relax ACM, 2018.

\end{thebibliography}
\end{document}